\documentclass[
  twocolumn,
  10pt,
  amsmath,
  amssymb,
  aps,
  prb,
  longbibliography,
  floatfix,
  superscriptaddress
]{revtex4-2}

\usepackage{graphicx}
\usepackage{dcolumn}
\usepackage{bm}
\usepackage{siunitx}
\usepackage{xcolor}
\usepackage[colorlinks=true,citecolor=blue,linkcolor=blue,urlcolor=blue]{hyperref}

\newcommand{\dd}{\mathrm{d}}
\newcommand{\ii}{\mathrm{i}}
\newcommand{\kb}{k_{\mathrm B}}
\newcommand{\eps}{\varepsilon}

\newcommand{\Phiiso}{\Phi_{\mathrm{iso}}}
\newcommand{\Phidim}{\Phi_{\mathrm{dim}}}
\newcommand{\Fdress}{F_{\mathrm{dress}}}
\newcommand{\Fbar}{\overline{F}_{\mathrm{dress}}}

\begin{document}

\title{Enhancement and Suppression of Thermal Emission from a Plasmonic Dimer}

\author{Victor Guillemot}
\affiliation{Institute of Industrial Science, The University of Tokyo, Tokyo, Japan}
\author{Roman Anufriev} 
\affiliation{Institute of Industrial Science, The University of Tokyo, Tokyo 153-8505, Japan}
\affiliation{LIMMS, CNRS-IIS IRL 2820, The University of Tokyo, Tokyo 153-8505, Japan}
\author{Michele Diego}
\affiliation{Univ. Lille, CNRS, Univ. Polytechnique Hauts-de-France, UMR 8520 – IEMN – Institut d’Electronique de Microélectronique et de Nanotechnologie, F-59650 Villeneuve- d’Ascq, France }
\author{Minggang Luo}
\affiliation{Institute of Industrial Science, The University of Tokyo, Tokyo, Japan}
\author{Jose Ordonez-Miranda}
\affiliation{Sorbonne Université, CNRS, Institut des Nanosciences de Paris, INSP, F-75005 Paris, France}
\author{Georges Hamaoui}
\affiliation{ESYCOM lab, Univ Gustave Eiffel, CNRS, ESYCOM, F-77454 Marne-la-Vallée, France}
\affiliation{LIMMS, CNRS-IIS IRL 2820, The University of Tokyo, Tokyo 153-8505, Japan}
\author{Sebastian Volz}
\affiliation{Laboratoire des Solides Irradiés Ecole Polytechnique, Institut Polytechnique de Paris - UMR CNRS 7642
- CEA/DRF/IRAMIS 28 route de Saclay, 91128 Palaiseau Cedex, France}
\author{Masahiro Nomura}
\affiliation{Institute of Industrial Science, The University of Tokyo, Tokyo, Japan}

\date{\today}

\begin{abstract}
Electromagnetic coupling can enhance or suppress the combined thermal emission
of nearby bodies relative to the sum of their individual contributions. The conditions driving this collective response are essential for controlling thermal radiation in closely spaced structures, but are poorly understood. 
Using full-wave fluctuational electrodynamics, here we study the thermal radiation from two coaxial gold cylinders as their separation decreases
from weak coupling to a metal--vacuum--metal cavity. As the gap narrows, plasmon resonances emerge, strengthen emission along the cylinder axis, and coexist with broadband suppression. A circular-cavity model connects these branches to gap-plasmon dispersion and lateral confinement.
We show a transition from integrated suppression to enhancement as
the resonances shift into the thermally occupied spectrum, depending on the radius and the gap distance. For a \SI{50}{\nano\meter} gap, the band-integrated net power emitted to the environment is about \SI{11}{\percent} lower than that of two isolated cylinders for a radius of \SI{0.5}{\micro\meter}, and about \SI{54}{\percent} higher for a radius of \SI{2}{\micro\meter}, for cylinders at \SI{301}{\kelvin} in an environment at \SI{300}{\kelvin}. The thermal benefit of cavity formation is therefore set
by the balance between resonant emission and the inhibited background,
rather than by peak enhancement alone.

\end{abstract}

\maketitle

\section{Introduction}
\label{sec:introduction}

The thermal emission of an object depends on its fluctuating currents and
its electromagnetic environment. Fluctuational
electrodynamics describes this dependence through the Green
tensor~\cite{polder_theory_1971,joulain_surface_2005}. Structuring the
available emission channels enables selective and directional
radiation~\cite{chu_controlling_2024,meinzer_plasmonic_2014,pendry_mimicking_2004,makhsiyan_shaping_2015,greffet_coherent_2002},
while multiple scattering between neighboring bodies modifies the emission
of each body~\cite{mulet_enhanced_2002,narayanaswamy_greens_2014,biehs_near-field_2021,greffet_unified_2026}. 

Coupling-induced enhancement and suppression have been studied theoretically in dipolar
many-body systems~\cite{song_many-body_2021,manjavacas_radiative_2012,herz_generalized_2022,nordlander_plasmon_2004}.
Extended objects support multiple spatial channels and require a full-wave
description. Recent studies have examined how electromagnetic coupling across the far-to-near field transition modifies the radiative flux~\cite{guillemot_nonmonotonic_2025,Guillemot2026NearFieldDressing}. This environmental dressing recalls Drexhage's experiment~\cite{drexhage_influence_1970}, with sources distributed through
dissipative bodies instead of localized in a single quantum emitter. Metal--insulator--metal (MIM) antennas provide a useful setting to study the collective response of extended bodies. In these structures,
gap plasmons propagate laterally and reflect at the edges to form resonances,
enabling selective absorption, field concentration, and infrared
emission~\cite{liu_taming_2011,liu_infrared_2010,costantini_plasmonic_2015,yang_ultrasmall_2012,petronijevic_active_2024,ma_broadband_2013,li_near-field_2018,chen_interference_2012}. In a metal--vacuum--metal cavity, thermal sources and dissipation reside in the metal, while the vacuum gap provides confinement. 

Here we investigate thermal radiation from a cavity formed by two micrometric gold cylinders using full-wave numerical calculations within fluctuational electrodynamics. We connect the
spectral, spatial, and angular response to the band-integrated net radiative
power for cylinders at \SI{301}{\kelvin} in an environment at \SI{300}{\kelvin}. The observable studied here is their net power emitted to the environment. While conventional studies of MIM thermal antennas exploit gap-plasmon resonances to tailor
spectral emission~\cite{liu_infrared_2010,costantini_plasmonic_2015}, here we quantify how cavity formation redistributes emission between
collective resonances and an inhibited broadband background, and how
this redistribution determines the net radiative-power balance relative
to two isolated emitters. Computations show that this balance can change sign with
size: resonant enhancement can coexist with reduced band-integrated
emission, whereas larger cylinders can emit more net power than the
isolated pair. The central result is thus a size-dependent competition
between resonant enhancement and broadband inhibition that determines
whether cavity formation increases or decreases the total
thermal emission.

Beyond this model system, understanding both the spectral and integrated
emission of individual microemitters can inform the design of thermal
metasurfaces for infrared sensing, radiative cooling, and thermophotovoltaic
energy conversion~\cite{chu_controlling_2024}. The spectral response sets the
frequencies at which radiation is delivered, whereas the integrated response
characterizes the overall radiative-power balance. An isolated emitter's
intrinsic response provides a reference for assessing how coupling modifies
these properties when emitters are brought together.

\begin{figure*}[t]
 \centering
 \includegraphics[width = 0.9\textwidth]{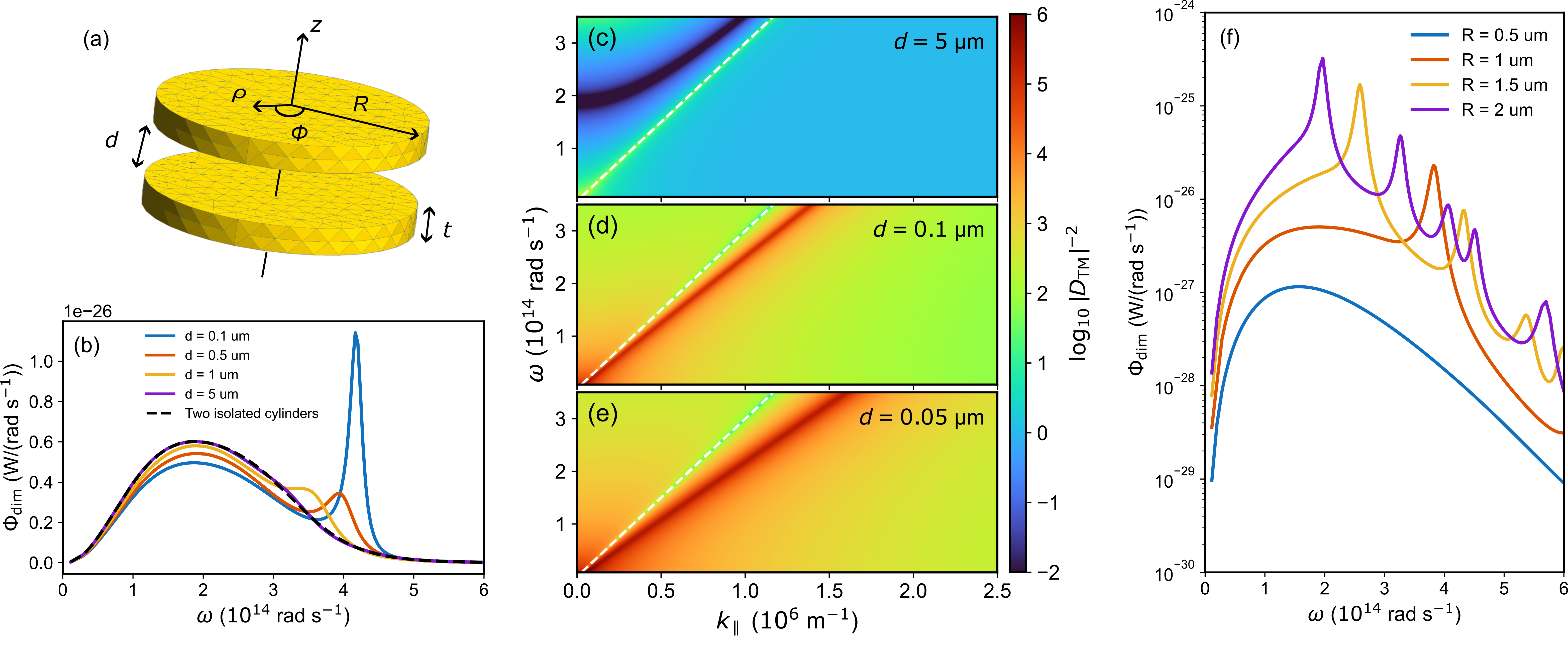}
 \caption{\label{fig:F1}
 Geometry and basic spectral response.
 (a) Two coaxial gold cylinders of radius $R$ and thickness $t$, separated by a vacuum gap $d$, with a representative surface mesh used in the BEM solver. 
 (b) Net radiative-power spectrum to the environment for $R=\SI{1}{\micro\meter}$, with both cylinders at \SI{301}{\kelvin} and the bath at \SI{300}{\kelvin}. The dashed curve represents two isolated cylinders ($d=\infty$) under the same conditions.
 (c)--(e) Planar Au--vacuum--Au dispersion for $d=\SI{5}{\micro\meter}$, $\SI{0.1}{\micro\meter}$, and $\SI{0.05}{\micro\meter}$, respectively. Colors show $\log_{10}|D_{\rm TM}|^{-2}$ on a common scale, where $D_{\rm TM}$ is the left-hand side of Eq.~\eqref{eq:mim_dispersion}. This dimensionless indicator locates the gap plasmon dispersion branch for the infinite radius case. Dashed lines denote the vacuum light line.
 (f) Net radiative-power spectrum to the environment at $T=\SI{301}{\kelvin}$ and $T_0=\SI{300}{\kelvin}$ for four radii $R$ and $d=\SI{50}{\nano\meter}$. Increasing the radius redshifts the fundamental resonance and brings higher frequency modes into the plotted band.}
\end{figure*}

\section{Geometry, formulation, and spectral response}
\label{sec:framework}

\begin{figure*}[t]
 \centering
 \includegraphics[width=0.90\linewidth]{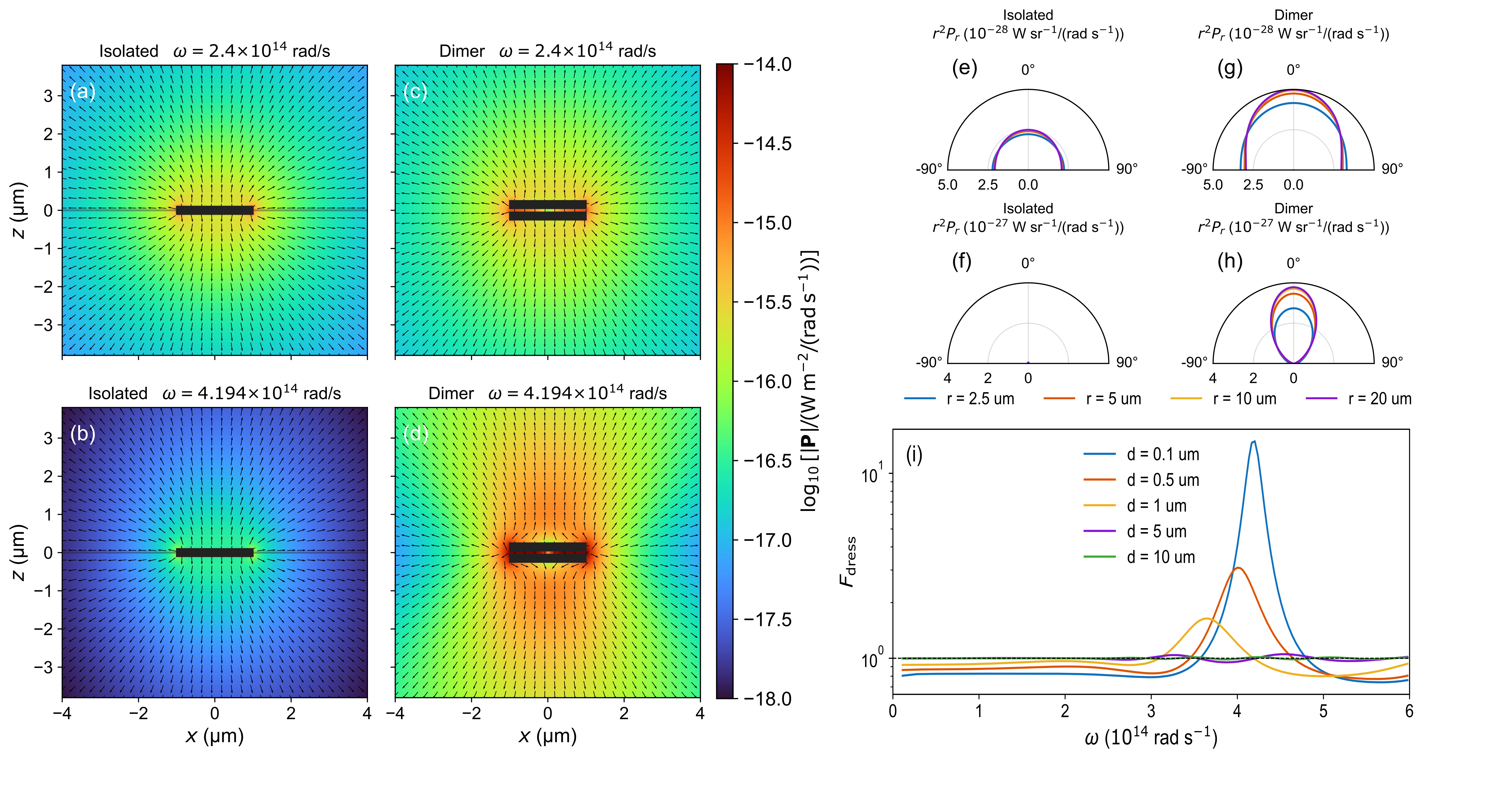}
 \caption{\label{fig:F2}
 Spatial and spectral redistribution of the emission. 
 (a)--(d) Poynting vector maps for the isolated cylinder (a,b) and the dimer (c,d), at two representative frequencies $\omega=\SI{2.4e14}{\radian\per\second}$ and $\omega=\SI{4.194e14}{\radian\per\second}$ for 
 $d=\SI{0.1}{\micro\meter}$. Colors show $\log_{10}[|\bm P|/(\si{\watt\per\square\meter\per\radian\second})]$ on one scale for all four maps. $\bm P$ is the net spectral Poynting vector in Eq.~\eqref{eq:poynting_maps}. Arrows give its direction in the $xz$ plane.
 (e)--(h) Angular intensities $r^2P_r$ in $\si{\watt\per\steradian\per\radian\second}$ for the isolated cylinder (e,f) and the dimer (g,h), evaluated at the same frequencies. Only the upper half-plane is shown, using mirror symmetry about the gap plane (or the isolated-cylinder midplane). Each frequency uses the same radial scale for its isolated and dimer panels. 
 (i) Dressing factor $F_\mathrm{dress}(\omega)$ for different gaps. The dashed line at unity denotes two independent cylinders. The cylinders are at \SI{301}{\kelvin} and the environment is at \SI{300}{\kelvin}.}
\end{figure*}

We consider two coaxial gold cylinders of radius $R$
and thickness $t=\SI{0.2}{\micro\meter}$ [Fig.~\ref{fig:F1}(a)].
Four radii $R=\SI{0.5}{\micro\meter}, \SI{1}{\micro\meter}, \SI{1.5}{\micro\meter}, \SI{2}{\micro\meter}$ are considered to span regimes from no resonance to multiple resonances within the thermally populated spectral window, as discussed below. The face-to-face vacuum gap $d$ ranges from \SI{50}{\nano\meter} to $\SI{20}{\micro\meter}$. Vacuum has a relative permittivity $\eps_\mathrm{vac}=1$ and contains no
thermal fluctuating currents. Gold is represented by the local Drude model~\cite{gold}
\begin{equation}
 \eps_m(\omega)=1-\frac{\omega_p^2}{\omega(\omega+\ii\gamma)},
 \label{eq:drude}
\end{equation}
where $\eps_m$ is the relative permittivity of gold, $\omega$ is the angular frequency, and $\ii$ is the imaginary unit. The plasma frequency is $\omega_p=\SI{1.37e16}{\radian\per\second}$ and
the Drude damping rate is $\gamma=\SI{5.32e13}{\radian\per\second}$.
We perform the full-wave numerical calculations using SCUFF-EM, an open-source electromagnetic solver based on the boundary-element method, with its fluctuating-surface-current formulation for thermal radiation~\cite{rodriguez_fluctuating-surface-current_2012}. The method represents the electromagnetic response using equivalent electric and magnetic currents on the discretized material interfaces. Thermal field correlations follow from the fluctuation--dissipation theorem, and the resulting fields determine the ensemble-averaged radiative fluxes, including multiple scattering between the cylinders.

Let $C_{ij}(\omega)$ denote the temperature-independent spectral transfer coefficient from fluctuating sources in object $i$ to receiving object $j$, including scattering by the complete structure. These coefficients are evaluated numerically with the boundary-element method. In the Landauer formalism~\cite{biehs_mesoscopic_2010}, the product of $C_{ij}$ with the thermal occupation factor provides power per unit angular frequency. The thermal occupation factor is given by the mean energy of a quantum harmonic oscillator at frequency $\omega$ and temperature $T$:
\begin{equation}
 \Theta(\omega,T)=\frac{\hbar\omega}{\exp(\hbar\omega/\kb T)-1}.
 \label{eq:planck}
\end{equation}
Here $\hbar$ is the reduced Planck constant and $\kb$ is the Boltzmann constant.
For two cylinders at a common temperature $T$ in an environment at $T_0$,
the net radiative-power spectrum to the environment is
\begin{equation}
 \Phidim(\omega,d,T,T_0)=[C_{10}(\omega,d)+C_{20}(\omega,d)]\Delta\Theta(\omega,T,T_0),
 \label{eq:net}
\end{equation}
where $\Delta\Theta(\omega,T,T_0)=\Theta(\omega,T)-\Theta(\omega,T_0)$.
Indices 1 and 2 label the cylinders, and 0 labels the environment.
The isolated-cylinder reference is
$\Phiiso(\omega,T,T_0)=C_{\mathrm{iso}}(\omega)\Delta\Theta(\omega,T,T_0)$.
Here $C_{\mathrm{iso}}$ is the transfer coefficient from a single isolated cylinder to the environment.
All radiative-power results use $T=\SI{301}{\kelvin}$ and $T_0=\SI{300}{\kelvin}$, with positive
power directed toward the environment. Both spectra vanish at $T=T_0$ and the 
net mutual radiative heat transfer vanishes because the two cylinders have equal temperatures.

We define the temperature difference as $\Delta T=T-T_0$ and denote by $P$ the net radiative power integrated over the simulated frequency band.
For $\Delta T=\SI{1}{\kelvin}$, $\Delta\Theta/\Delta T$ agrees with $\left.\partial_T\Theta\right|_{\SI{300}{\kelvin}}$ to within a few percent over the plotted frequency band, justifying this temperature difference for interpreting $P/\Delta T$ as a radiative conductance.

Table~\ref{tab:convergence} reports the completed mesh-refinement study
at $R=\SI{2}{\micro\meter}$ and at the smallest gap, $d=\SI{50}{\nano\meter}$.
The finest mesh ($h=\SI{0.05}{\micro\meter}$) covers all 176 sampled frequencies.
Refinement from $h=\SI{0.075}{\micro\meter}$ to $\SI{0.05}{\micro\meter}$ changes the thermally weighted spectrum by \SI{1.09}{\percent} in the integrated absolute-difference measure and the band-integrated power by \SI{0.32}{\percent}, supporting mesh convergence for this configuration.
Band-integrated powers use the same frequency interval for the dimer and the isolated reference, with trapezoidal integration.
\begin{table}[t]
\caption{Mesh convergence for $R=\SI{2}{\micro\meter}$ and $d=\SI{50}{\nano\meter}$.
$N_\triangle$ counts surface triangles across both cylinders, $h$ is the
nominal mesh size, specifying the target mean triangle-edge length.
Elapsed wall times are taken from the 32-thread numerical-solver logs and
include the frequency sweep. Errors use the finest available mesh as reference and the exact \SI{301}{\kelvin}/\SI{300}{\kelvin} thermal weight over the common band
$[\SI{0.1125e14}{\radian\per\second},\SI{5.9763e14}{\radian\per\second}]$.
The zero entries define the reference and do not imply zero absolute error.}
\label{tab:convergence}
\begin{ruledtabular}
\begin{tabular}{ccccc}
$h$ (\si{\micro\meter}) & $N_\triangle$ & Time (\si{\hour}) & $\epsilon_{\rm spec}$ (\si{\percent}) & $\epsilon_P$ (\si{\percent}) \\
0.05 & 14882 & $\geq 89.64$ & 0.00 & 0.00 \\
0.075 & 6448 & 17.18 & 1.09 & 0.32 \\
0.1 & 3680 & 4.91 & 2.78 & 1.05 \\
0.2 & 1104 & 1.61 & 7.48 & 3.11 \\
0.4 & 384 & 0.40 & 19.33 & 7.85 \\
0.8 & 88 & 0.10 & 52.26 & 29.14 \\

\end{tabular}
\end{ruledtabular}
\end{table}
The relative errors $\epsilon_{\rm spec}$ and $\epsilon_P$ measure, respectively, the frequency-integrated absolute spectral difference and the absolute difference in band-integrated power, both normalized by the reference band-integrated power and expressed as percentages.

Figure~\ref{fig:F1}(b) represents the net radiative-power spectrum to the environment for $R=\SI{1}{\micro\meter}$, with both cylinders at \SI{301}{\kelvin} and the bath at \SI{300}{\kelvin}. The dashed curve represents the case of two isolated cylinders ($d=\infty$) under the same conditions (i.e. twice the emission of a single cylinder). At short gaps, a collective resonance near
$\SI{4.2e14}{\radian\per\second}$ coexists with off-resonant inhibition. Over this range of gaps, both the resonance frequency and the quality factor increase as the gap decreases.
The planar dispersion maps [Figs.~\ref{fig:F1}(c)--\ref{fig:F1}(e)], computed analytically for two facing, infinite gold surfaces, illustrate increasing confinement as the gap decreases. The bound branch corresponding to the gap plasmon modes moves away from the vacuum light line as the gap is reduced. In the case of the antenna, the field is confined along $z$ but propagates laterally, where the finite radius $R$ imposes a resonance
condition. Figure~\ref{fig:F1}(f), plotted on a logarithmic scale, shows the effect of increasing the radius, which redshifts the fundamental resonance and introduces higher-order resonances into the available band. Increasing the radius also increases the overall emission as the cylinder surface area increases. For the smallest radius $R= \SI{0.5}{\micro\meter}$ (blue curve), the fundamental resonance is above the thermally weighted spectrum and no resonance appears within this spectral range. Increasing the radius brings additional modes into the simulated band. 

\section{Spatial and angular signatures}
\label{sec:spatial}

Figure~\ref{fig:F2} compares an isolated cylinder with a dimer at
$d=\SI{100}{\nano\meter}$, for $R=\SI{1}{\micro\meter}$. The maps show the
net spectral Poynting vector obtained from the numerically evaluated, spatially resolved
response coefficients $\bm p_i$:
\begin{equation}
 \bm P(\bm r,\omega)=\Delta\Theta(\omega;T,T_0)
 \sum_{i=1}^{2}\bm p_i(\bm r,\omega),
 \label{eq:poynting_maps}
\end{equation}
where $\bm p_i$ describes the ensemble-averaged response to fluctuating
currents in cylinder $i$, including scattering by the complete structure.
For the isolated cylinder, the sum contains a single source contribution.
The thermal flux difference accounts for the environment at $T_0$ relative to
the cylinders at $T$. The spatial coefficients are obtained by summing
Cartesian source contributions before
taking the vector magnitude. 
Figures~\ref{fig:F2}(a)--(d) show a broader spatial emission pattern off resonance and an enhanced axial response at resonance. Colors show the magnitude
$|\bm P|=\sqrt{P_x^2+P_y^2+P_z^2}$ on a common logarithmic scale and 
arrows indicate the direction of the Poynting vector in the $y=0$ plane.
At resonance, the dimer strongly enhances the response around the cavity and along
its axis. The angular plots in Figs.~\ref{fig:F2}(e)--(h) show radiation patterns evaluated on a circular grid of radius $r$ in the $xz$ plane. The quantity $r^2P_r$ is shown, where
$P_r=P_x\sin\theta+P_z\cos\theta$ and $\theta$ is measured from $+z$.
The resonant dimer enhances the axial emission of an isolated cylinder, which already has a directional pattern at this frequency. The axial intensity increase and the angle-integrated enhancement are distinct observables. At $\omega=\SI{4.194e14}{\radian\per\second}$, direct integration of the angular data gives a dimer directivity
$\mathcal D=4\pi\max(r^2P_r)/\int r^2P_r\,\dd\Omega$ of 4.22
at $r=\SI{20}{\micro\meter}$, with a full angular width at half of the axial
maximum of \SI{69.0}{\degree}. Here $\dd\Omega$ is the solid-angle element and the integral extends over the full sphere. The isolated-cylinder directivity at the same
radius is 1.61.
Integrating the angular flux at $\SI{20}{\micro\meter}$ agrees with the
interpolated total transfer coefficient $C_{10}+C_{20}$ within \SI{0.07}{\percent} off
resonance and \SI{2.15}{\percent} at resonance for the dimer.

The angle-integrated spectral enhancement is measured by the defined dressing factor $F_\mathrm{dress}(\omega,d)$, which compares the dimer emission to twice the isolated-cylinder reference at a given frequency:
\begin{equation}
  F_\mathrm{dress}(\omega,d)
  = \frac{\Phi_\mathrm{dim}(\omega,d)}{2\Phi_\mathrm{iso}(\omega)}
  = \frac{C_{10}(\omega,d)+C_{20}(\omega,d)}{2C_\mathrm{iso}(\omega)}.
  \label{eqn:Dress}
\end{equation}

Figure~\ref{fig:F2}(i) shows more than tenfold resonant enhancement
and off-resonant suppression. The dashed line at $F_\mathrm{dress}(\omega,d)=1$ separates enhancement from suppression. The cavity feature emerges relative to the
isolated-cylinder spectrum, in contrast to dimers whose constituents already
support infrared phonon-polaritonic modes~\cite{tang_corner-_2024,zeng_mode_2025,OrdonezMiranda2026Enhanced}.
The transfer coefficients $C_{ij}$ and the local density of states both
depend on the Green tensor, but probe different observables: transfer between
source and receiving channels versus a local response~\cite{carminati_electromagnetic_2015,joulain_definition_2003}.

\section{Analytical circular-cavity model}
\label{sec:analytical}

\begin{figure*}[t]
 \centering
 \includegraphics[width=0.90\linewidth]{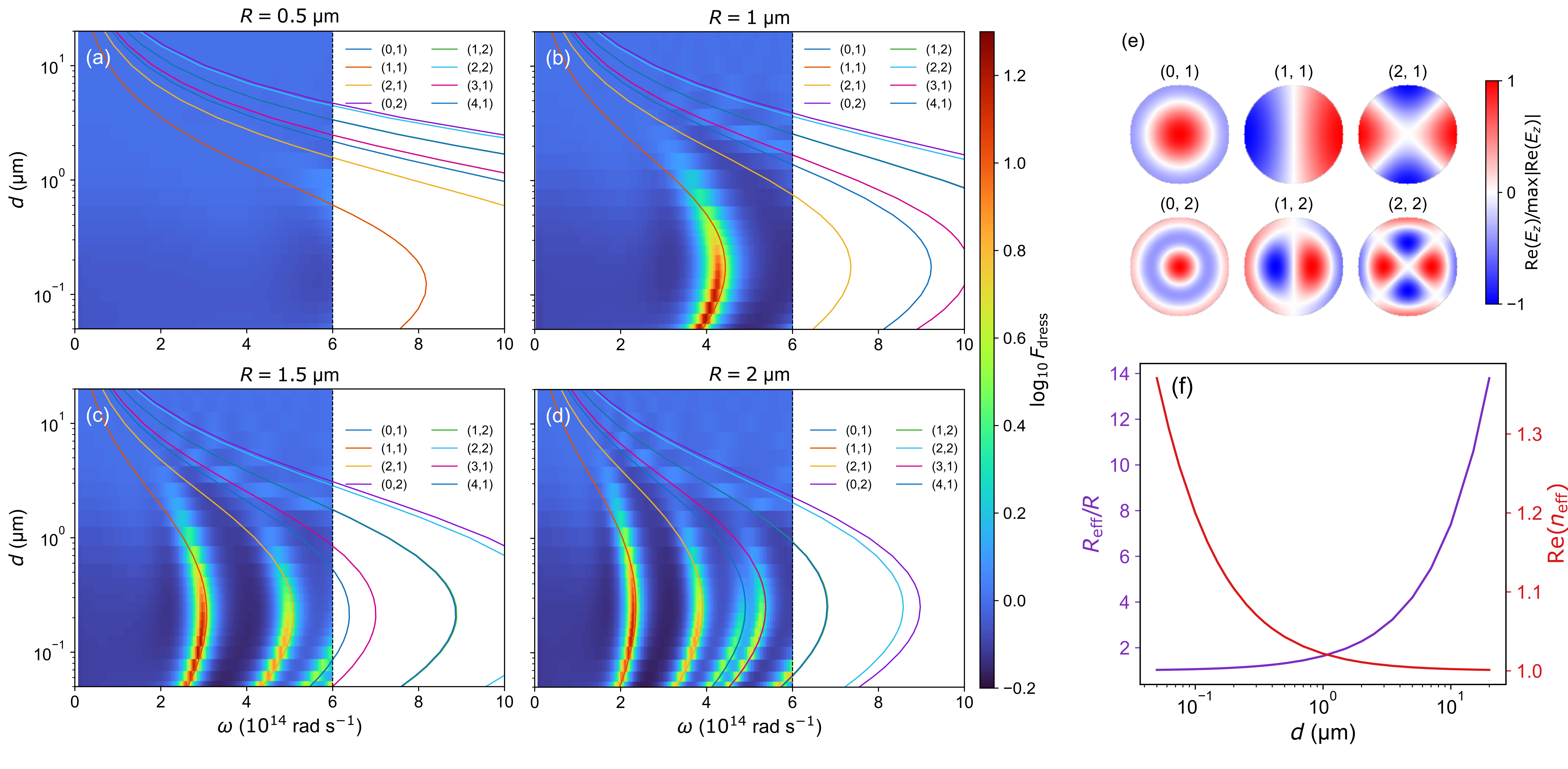}
 \caption{\label{fig:F3}
 Spectral and spatial maps for four radii with analytical mode trajectories.
 (a)--(d) Numerically calculated dressing factor as a function of frequency and gap.
 Panels (a)--(d) correspond to $R=\SI{0.5}{\micro\meter}$, $\SI{1}{\micro\meter}$, $\SI{1.5}{\micro\meter}$, and $\SI{2}{\micro\meter}$, respectively. Color represents $\log_{10}F_{\mathrm{dress}}$ on the same linear color scale in all four panels, negative values indicate inhibition and positive values indicate enhancement. Curves denote eight analytical circular-cavity branches, including the additional $(3,1)$ and $(4,1)$ modes. The colored maps extend to $\SI{6e14}{\radian\per\second}$, curves in the white region at higher frequencies are analytical predictions without a numerical comparison.
 (e) Normalized electric-field profiles of the six $(m,n)$ modes,
 $\widetilde E_z=\operatorname{Re}E_z/\max|\operatorname{Re}E_z|$, at
 $d=\SI{0.1}{\micro\meter}$ and the corresponding model frequencies.
 In-plane coordinates are normalized by the physical radius.
 (f) Effective-radius ratio $R_{\mathrm{eff}}/R$ (purple, left axis) and real effective index (red, right axis) versus gap for $R=\SI{1}{\micro\meter}$.
 The index is evaluated along the fundamental $(m,n)=(1,1)$ branch and 
 the radius ratio follows Eq.~\eqref{eq:Reff}.}
\end{figure*}

Cavity modes in circular patch nanoantennas arise from the propagation and
reflection of gap plasmons at the antenna boundary, which produce standing
waves within the dielectric gap. Following the cavity descriptions developed
by Filter \textit{et al.}~\cite{filter_circular_2012} and Minkowski
\textit{et al.}~\cite{minkowski_resonant_2014}, we combine the planar MIM
dispersion with an approximate lateral confinement condition to calculate
resonance frequencies and normalized field profiles. Here, the edge response
is represented by an effective-radius correction rather than an explicit
calculation of the complex reflection coefficient.

For a planar MIM waveguide, we define
\begin{equation}
 k_0=\frac{\omega}{c},\qquad
 \kappa_\mathrm{vac}=\sqrt{\beta^2- k_0^2},\qquad
 \kappa_\mathrm{m}=\sqrt{\beta^2-\eps_\mathrm{m}k_0^2}.
\end{equation}
Here $c$ and $k_0$ are speed of light and wavevector in vacuum respectively, while $\kappa_\mathrm{vac}$ and $\kappa_\mathrm{m}$ are the transverse wavevector in the vacuum gap and metal, respectively. The transverse-magnetic (TM) modes that decay evanescently across the interface, satisfy the following dispersion relation~\cite{Economou}
\begin{equation}
 \tanh\!\left(\frac{\kappa_\mathrm{vac} d}{2}\right)
 +\frac{\kappa_\mathrm{m}}{\eps_\mathrm{m}\kappa_\mathrm{vac}}=0,
 \label{eq:mim_dispersion}
\end{equation}
which we solve for the complex in-plane wavevector
$\beta=\beta'+\ii\beta''$. The corresponding effective index is
\begin{equation}
 n_{\mathrm{eff}}(\omega,d)=\frac{\beta(\omega,d)}{k_0}.
 \label{eq:neff}
\end{equation}
Its real part determines the phase velocity, while its imaginary part
describes propagation loss. Reducing the gap increases the phase index
in the confined regime and thus shortens the lateral plasmon wavelength (see Figs.~\ref{fig:F1}(c)--\ref{fig:F1}(e)).

Within the cavity approximation, we separate the radial, azimuthal, and
axial dependences of the normal electric-field component. Using cylindrical
coordinates centered on the gap leads to
\begin{equation}
 E_z(\rho,\phi,z)=E_0J_m(\beta\rho)
 \cos(m\phi+\varphi)
 \frac{\cosh(\kappa_\mathrm{vac}z)}{\cosh(\kappa_\mathrm{vac}d/2)},
 \label{eq:Ez_mode}
\end{equation}
Here $E_z$ is the complex electric-field component along the common cylinder axis and $E_0$ is an arbitrary complex amplitude. The cylindrical coordinates $(\rho,\phi,z)$ are defined in Fig.~\ref{fig:F1}(a): $\rho$ is the radial distance from the axis, $\phi$ is the azimuthal angle measured in the transverse plane from a fixed reference direction, and $z$ is the axial coordinate, with $z=0$ at the gap midplane and the facing metal surfaces at $z=\pm d/2$. The function $J_m$ is the Bessel function of the first kind of integer order $m\geq0$, which labels the azimuthal mode dependence. The constant angular phase $\varphi$ selects the in-plane orientation of the standing-wave pattern, for $m>0$, an angular rotation $\phi_0$ corresponds to $\varphi=-m\phi_0$. The complex propagation constant $\beta$ describes lateral propagation and loss, while $\kappa_{\mathrm{vac}}$ sets the axial field variation across the vacuum gap of width $d$, the denominator normalizes the axial factor to unity at the two facing surfaces. The displayed field profiles are the real parts of these complex fields. An approximate Neumann
condition at an effective boundary leads to the real-frequency phase condition
\begin{equation}
 \operatorname{Re}\beta(\omega_{mn},d)R_{\mathrm{eff}}(d)=u'_{mn},
 \label{eq:resonance_condition}
\end{equation}
Here $\omega_{mn}$ is the real resonance angular frequency of the mode $(m,n)$, $n\geq1$ is its radial index, and $R_{\mathrm{eff}}$ is the effective cavity radius. The quantity $u'_{mn}$ is the $n$th positive root of $J_m'$, where the prime denotes differentiation with respect to the Bessel-function argument. The root at $u = 0$ is excluded, $(0,1)$ therefore, denotes the first nonzero radial solution for $m=0$. 
We model the edge correction with an effective radius that increases linearly with gap:
\begin{equation}
 R_{\mathrm{eff}}(d)=R+\alpha d.
 \label{eq:Reff}
\end{equation}
A single dimensionless boundary coefficient, $\alpha$, is used for all radii and modes.
To calibrate $\alpha$, we minimize the mean squared relative frequency
error between the resonance frequencies predicted by Eq.~\eqref{eq:resonance_condition} and the peak frequencies of the numerically calculated transfer coefficient between the dimer and the environment ($C_{10} + C_{20}$). The calibration uses radii with resolved peaks and applies no thermal weighting. This procedure yields
$\alpha=0.64$, with a root-mean-square (RMS) relative frequency error of $\SI{3.9}{\percent}$.
All analytical curves and mode profiles use this optimized value.

Figures~\ref{fig:F3}(a)--\ref{fig:F3}(d) show
$\log_{10}F_{\mathrm{dress}}$ where positive and negative values indicate
enhancement and inhibition, respectively. The colored maps cover the simulated frequency band, only the analytical trajectories extend beyond its upper boundary. As the gap decreases, resonant branches emerge, blueshift, then redshift in the confined regime.
The inter-resonance background is suppressed, while the large-gap response
approaches the independent-emitter limit ($F_{\mathrm{dress}}(\omega,d)=1$).

The superimposed curves show eight analytical branches:
$(m,n)=(0,1)$, $(1,1)$, $(2,1)$, $(0,2)$, $(1,2)$, $(2,2)$, $(3,1)$, and $(4,1)$. Only branches inside the simulated band can be compared with the numerically calculated
spectra. In particular, the smallest-radius case does
not test the branches lying above the simulated band. 
Figure~\ref{fig:F3}(e) shows the corresponding mid-gap field profiles at
$d=\SI{100}{\nano\meter}$, each evaluated at its modelled frequency and displayed as
$\operatorname{Re}E_z/\max|\operatorname{Re}E_z|$. The $(1,1)$ branch is the fundamental
nontrivial mode of this circular-cavity model. Its out-of-plane electric
field has dipolar azimuthal symmetry, with two opposite-sign lobes and
two degenerate orthogonal orientations. Agreement between a spectral peak
and a modelled frequency provides a candidate mode assignment while the field
symmetry offers a separate test of that assignment.

To test additional features in Fig.~\ref{fig:F3}(d), we also solve the
same phase condition, without recalibrating $\alpha$, for $(m,n)=(3,1)$
and $(4,1)$. At $R=\SI{2}{\micro\meter}$ and $d=\SI{50}{\nano\meter}$, their predicted
frequencies are $\SI{4.522e14}{\radian\per\second}$ and $\SI{5.723e14}{\radian\per\second}$.
The transfer coefficient to the environment has local maxima at $\SI{4.509e14}{\radian\per\second}$ and
$\SI{5.702e14}{\radian\per\second}$, respectively. Thus the extra
peak near $\SI{4.5e14}{\radian\per\second}$ is consistent with a
$(3,1)$ assignment (\SI{0.3}{\percent} frequency detuning at this gap).
The $(4,1)$ frequency is nearly degenerate with the existing $(1,2)$
prediction, $\SI{5.738e14}{\radian\per\second}$. These are candidate assignments
pending a comparison of full-wave field symmetries.

The spectral turning point reflects a competition between propagation and
the effective boundary. Differentiating Eq.~\eqref{eq:resonance_condition}
along a phase branch yields
\begin{equation}
 \frac{\dd\omega_{mn}}{\dd d}=-
 \frac{R_{\mathrm{eff}}\,\partial_d\operatorname{Re}\beta
 +\operatorname{Re}\beta\,\partial_dR_{\mathrm{eff}}}
 {R_{\mathrm{eff}}\,\partial_\omega\operatorname{Re}\beta}.
 \label{eq:shift_decomposition}
\end{equation}
Provided the denominator is nonzero, a frequency extremum with respect to gap, $\dd\omega_{mn}/\dd d=0$, satisfies
\begin{equation}
 R_{\mathrm{eff}}\,\partial_d\operatorname{Re}\beta
 +\alpha\operatorname{Re}\beta=0.
 \label{eq:turning_point}
\end{equation}
The position of the frequency maximum depends on radius and mode through the
dispersion and its derivatives.

When the phase index varies weakly with gap, reducing $d$ decreases
$R_{\mathrm{eff}}$ and tends to shift the resonance upward. At smaller gaps,
the increase in the phase index can dominate and shift the resonance
downward. The frequency maximum marks the balance of these contributions.
Figure~\ref{fig:F3}(f) illustrates the effective radius and the phase index
along the fundamental $(1,1)$ branch for $R=\SI{1}{\micro\meter}$. The evolution of these two quantities with gap explains the successive blueshift and redshift of the resonance. 
The model assumes metal layers that are semi-infinite along $z$.
At $\omega=\SI{4e14}{\radian\per\second}$, the Drude model gives
an amplitude skin depth $\delta=[\operatorname{Im}(k_0\sqrt{\eps_m})]^{-1}
\simeq\SI{22}{\nano\meter}$, so that $t\simeq9\delta$.
The cavity approximation is most appropriate for \(d\ll R\). At gaps comparable to or larger than the radius, its predicted branches should be interpreted cautiously, whereas the full-wave calculation includes the radiative interactions between the finite cylinders.

\begin{figure*}[t]
 \centering
 \includegraphics[width=0.82\linewidth]{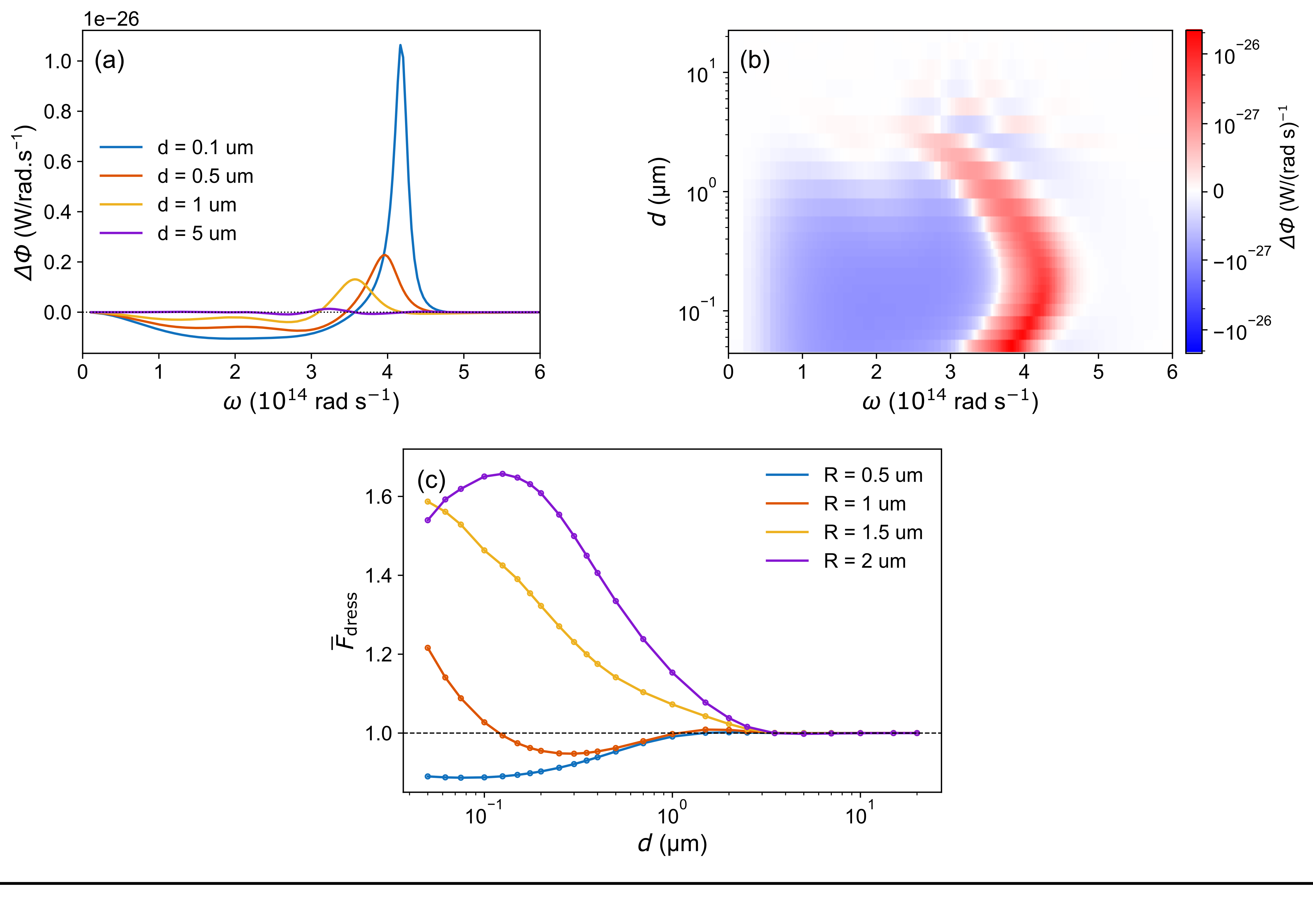}
 \caption{\label{fig:F4}
 From spectral redistribution to net radiative power, for cylinders at
 $T=\SI{301}{\kelvin}$ in an environment at $T_0=\SI{300}{\kelvin}$.
 (a) Excess net power spectrum $\Delta\Phi=\Phidim-2\Phiiso$
 for $R=\SI{1}{\micro\meter}$ at selected gaps.
 Positive and negative values denote enhancement and suppression relative
 to two isolated cylinders under the same thermal conditions.
 (b) Gap dependence of the same excess spectrum. Red and blue regions identify enhanced and suppressed spectral contributions, respectively. White regions indicates where the excess spectrum is close to zero (even when the dressing factor is significant). Colorbar labels give $\Delta\Phi$ in $\si{\watt\per\radian\second}$.
 (c) Band-integrated dressing factor for
 $R=\SI{0.5}{\micro\meter}, \SI{1}{\micro\meter}, \SI{1.5}{\micro\meter}, \SI{2}{\micro\meter}$ as a function of gap. Each curve is normalized by twice
 the isolated-cylinder net power for the same radius, temperatures, and
 integration band. The dashed line at unity marks the independent-emitter
 limit. }
\end{figure*}

\section{Band-integrated emission to the environment}
\label{sec:integrated}

Figure~\ref{fig:F4} connects the spectral changes induced by coupling to
their net radiative-power balance. Throughout this section, the cylinders
are at $T=\SI{301}{\kelvin}$ and the environment is at $T_0=\SI{300}{\kelvin}$. For
$R=\SI{1}{\micro\meter}$, Figs.~\ref{fig:F4}(a) and \ref{fig:F4}(b) show
the excess spectrum
\begin{equation}
 \Delta\Phi(\omega,d;T,T_0)
 =\Phidim(\omega,d;T,T_0)-2\Phiiso(\omega;T,T_0).
 \label{eq:excess_spectrum}
\end{equation}
This difference isolates the effect of coupling from the emission of two
independent cylinders. In contrast to the ratio, $F_\mathrm{dress}$ (Eq.~\ref{eqn:Dress}), in which the thermal weighting factors cancel out, here a negative value indicates reduced outward power relative to this reference.

At short gaps, a narrow positive resonant contribution coexists with a
broader negative contribution at lower frequencies. The map in
Fig.~\ref{fig:F4}(b) follows their evolution with separation: the red
resonant band shifts with gap, while the blue region reveals suppression
of the background. Both contributions weaken as the cylinders become
independent. At high frequencies, the thermal difference
$\Delta\Theta$ suppresses the net spectrum (white regions in Fig.~\ref{fig:F4}(b)). At low frequencies, $\Delta\Theta\to\kb(T-T_0)$ remains finite and the small excess therefore reflects the electromagnetic response and its difference from the isolated reference.

The net power over the simulated band is
\begin{equation}
 P_{\mathrm{dim}}(d;T,T_0)=
 \int_{\omega_{\min}}^{\omega_{\max}}
 \Phidim(\omega,d;T,T_0)\,\dd\omega,
 \label{eq:integrated_power}
\end{equation}
with $P_{\mathrm{iso}}$ evaluated over the same interval. A large resonant peak can therefore coexist
with reduced integrated power if its contribution is outweighed by
broadband suppression.

Figure~\ref{fig:F4}(c) summarizes this balance through the band-integrated 
dressing factor,
\begin{equation}
 \Fbar(d;T,T_0)=
 \frac{P_{\mathrm{dim}}(d;T,T_0)}{2P_{\mathrm{iso}}(T,T_0)}.
 \label{eq:integrated_dress}
\end{equation}
The connection to the spectral dressing factor is explicit in
\begin{equation}
 \Fbar(d;T,T_0)=
 \frac{\displaystyle\int_{\omega_{\min}}^{\omega_{\max}}
 \Fdress(\omega,d)\Phiiso(\omega;T,T_0)\,\dd\omega}
 {\displaystyle\int_{\omega_{\min}}^{\omega_{\max}}
 \Phiiso(\omega;T,T_0)\,\dd\omega}.
 \label{eq:weighted_dressing}
\end{equation}

Values above and below unity indicate enhanced and suppressed emission,
respectively. Figure~\ref{fig:F4}(c) shows that the integrated response
varies nonmonotonically with separation for several radii, with the
direction and magnitude of the change depending on the geometry. Thus,
decreasing the gap does not necessarily increase or decrease the net
radiative power: the outcome depends on the balance between resonant
enhancement and broadband suppression. More specifically, the figure shows suppression for
$R=\SI{0.5}{\micro\meter}$ at short gaps, with a dressing factor of about
0.89. The $R=\SI{1}{\micro\meter}$ curve is also nonmonotonic: it decreases
from about 1.22 at $d=\SI{50}{\nano\meter}$ to a minimum of about 0.95 at an
intermediate submicrometer gap, then returns toward unity as separation
increases. Thus this radius displays suppression over a substantial gap
interval, despite its strong spectral resonance. For $R=\SI{1.5}{\micro\meter}$,
the factor is about 1.59 at the smallest gap and decreases toward unity
with increasing separation. For $R=\SI{2}{\micro\meter}$, it is about 1.54
at the smallest gap and reaches a maximum of about 1.66 at a
finite gap near $\SI{0.1}{\micro\meter}$, before decreasing toward unity.
These approximate values describe the curves displayed in
Fig.~\ref{fig:F4}(c). The largest radius therefore illustrates that
reducing the gap beyond the optimum does not necessarily increase the integrated
response. All four curves approach the independent-emitter limit at large
separations.

The integrated factor is therefore a weighted average of the spectral
factor, with weight $\Phiiso=C_{\mathrm{iso}}\Delta\Theta$.
Consequently, strong optical dressing at a particular frequency can lead
to only a moderate change in the integrated thermal power if that
frequency contributes little to the thermally weighted spectrum, or if
enhancement is offset by suppression elsewhere.
Changing radius modifies both the collective resonances and the
isolated-cylinder reference. Changing gap redistributes enhancement and
suppression across that spectrum. Resonance frequencies, linewidths,
amplitudes, and the inhibited background all enter the resulting balance.
Figure~\ref{fig:F4} thus establishes the relevant criterion for enhanced
collective thermal radiation: the resonances must increase the net
spectral area over the thermally weighted band, not merely the peak height.

\section{Discussion}
\label{sec:discussion}

The principal result is that cavity formation does not by itself guarantee
enhanced thermal radiation to the environment. The relevant comparison is
with two isolated cylinders of the same radius, rather than with a fixed
geometric emitting area. Equation~\eqref{eq:weighted_dressing} shows that
the integrated response depends on the overlap of the collective spectrum
with the isolated-cylinder spectrum and the thermal occupation difference.
Consequently, neither a large spectral dressing factor nor enhanced axial
intensity is sufficient to establish an increase in net radiative power.
The radius and gap control this balance by shifting the resonant branches
and modifying their contribution relative to the inhibited background.

The circular-cavity model identifies gap-plasmon propagation and lateral
confinement as the ingredients governing the resonance trajectories.
Its phase condition also separates the effects of the gap-dependent
propagation constant and the effective boundary, providing an
interpretation of the successive blueshift and redshift. However, the
effective-radius correction is a calibrated description of the edge
response. The model does not calculate the radiative coupling strengths,
resonance linewidths, or broadband background, and therefore does not
independently predict the integrated enhancement. Those quantities are
determined here by the full-wave calculation. Likewise, agreement in
resonance frequency supports a mode assignment but does not replace a
comparison of full-wave field symmetries, particularly for nearly
degenerate branches.

Away from resonance, mutual screening offers a qualitative interpretation
of the reduced emission to the environment. This interpretation should
not be reduced to a loss of exposed surface area: diffraction,
interference, and multiple scattering determine the electromagnetic
response of the finite cylinders. The present calculations establish the
suppressed background but do not separately quantify these contributions.
Because both cylinders have the same temperature, the observed change
concerns their collective exchange with the environment, while their net
mutual heat transfer remains zero.

The quantitative balance reported here applies to the local Drude
description, the sampled geometries, the simulated frequency band, and
the specified temperatures. The phase model is most appropriate for
$d\ll R$; its large-gap trajectories should be treated as extrapolations
of the effective-boundary approximation. In addition, although the
spectral dressing factor is independent of the thermal occupation
difference within the fixed material model, its integrated counterpart
depends on temperature through the spectral weight. A geometry that
enhances the net power at one temperature need not retain the same
advantage at another. These results therefore motivate optimizing the
thermally weighted spectral balance, with the resonance position,
linewidth, and background considered together.

\section{Conclusion}
We have shown that electromagnetic coupling between two coaxial gold cylinders can either enhance or suppress their band-integrated net thermal emission to the environment relative to two isolated cylinders. The balance between collective resonances and broadband inhibition depends on cylinder radius and separation, yielding a nonmonotonic gap dependence in some geometries. The circular-cavity model interprets the resonance trajectories through gap-plasmon dispersion and lateral confinement, while the full-wave calculations determine the resulting radiative-power balance. These findings identify the thermally weighted spectral balance as the relevant criterion for optimizing collective thermal emission.

\begin{acknowledgments}
\textit{Funding:} This work was supported by the Japan Society for the Promotion of Science through JSPS KAKENHI grant number 26KF0129 (M. Nomura).
\textit{Competing interests:} The authors declare that they have no competing interests. 
\end{acknowledgments}

\bibliographystyle{apsrev4-2}
\bibliography{Papier_3}

\end{document}